\documentclass[a4paper]{article}
\usepackage[pdftex]{graphicx}
\usepackage{hhline}
\usepackage{tabularx}
\usepackage{INTERSPEECH2022}
\usepackage{cite}

\newcolumntype{Y}{&gt;{\centering\arraybackslash}X} 

\title{Speaker consistency loss and step-wise optimization for semi-supervised joint training of TTS and ASR using unpaired text data}
\name{Naoki Makishima, Satoshi Suzuki, Atsushi Ando, Ryo Masumura}
\address{
  NTT Corporation, Japan}
\email{naoki.makishima.fx@hco.ntt.co.jp}
\begin{document}

\maketitle
\begin{abstract}
In this paper, we investigate the semi-supervised joint training of text to speech (TTS) and automatic speech recognition (ASR), where a small amount of paired data and a large amount of unpaired text data are available.
Conventional studies form a cycle called the TTS-ASR pipeline, where the multi-speaker TTS model synthesizes speech from text with a reference speech and the ASR model reconstructs the text from the synthesized speech, after which both models are trained with a cycle-consistency loss.
However, the synthesized speech does not reflect the speaker characteristics of the reference speech and the synthesized speech becomes overly easy for the ASR model to recognize after training.
This not only decreases the TTS model quality but also limits the ASR model improvement.
To solve this problem, we propose improving the cycle-consistency-based training with a speaker consistency loss and step-wise optimization.
The speaker consistency loss brings the speaker characteristics of the synthesized speech closer to that of the reference speech.
In the step-wise optimization, we first freeze the parameter of the TTS model before both models are trained to avoid over-adaptation of the TTS model to the ASR model.
Experimental results demonstrate the efficacy of the proposed method.
\end{abstract}

\section{Introduction\label{sec:intro}}
Text to speech (TTS) and automatic speech recognition (ASR) perform inverse tasks of each other.
Although these two models were developed independently, several recent studies have investigated learning both the TTS model and ASR model simultaneously~\cite{ATjandra20_MachineSpeechChain,SKarita19_SharedTTSASR,THori19_CycleConsistencyTraining,YRen19_almostunsupervised,JXu20_LRSpeech,MBaskar21_EAT}.
One of the application in these studies is semi-supervised joint training, which creates models with a small amount of paired (speech and text) data and a large amount of unpaired text data.
Compared to other semi-supervised training methods of TTS~\cite{YChung19_semisupervisedTTS_embed1,Hwang21_TTSbyTTS,Hwang21_PnGBERT} and ASR~\cite{AKannan18_shallowfusion,ASriram18_coldfusion,ARosenberg19_ASRaugmentedwithTTS}, the advantage of semi-supervised joint training is that knowledge obtained by one model spreads to the other model, which is closely related to the speech chain mechanism of human communication~\cite{denes1993speech}.
Thus, semi-supervised joint training of TTS and ASR is a promising approach for applications that learn while speaking and listening like a human.
In this paper, we focus on semi-supervised joint training utilizing paired data and unpaired text data.

Various studies have investigated the semi-supervised joint training of the TTS model and ASR model~\cite{ATjandra20_MachineSpeechChain,SKarita19_SharedTTSASR,THori19_CycleConsistencyTraining,YRen19_almostunsupervised,JXu20_LRSpeech,MBaskar21_EAT}.
When using unpaired text data for training, a cycle called the TTS-ASR pipeline is utilized, where the multi-speaker TTS model synthesizes speech from the text with a randomly chosen reference speech and the ASR model transcribes the synthesized speech to reconstruct the source text.
The conventional studies use the cycle-consistency loss~\cite{DHe16_CCMT,JZhu17_CCCV} that evaluates the distance between the source text and the reconstructed text to update the TTS model and ASR model in the pipeline.
This training improves both models; the TTS model learns to correct the incorrect pronunciations so that the ASR model can accurately reconstruct the source text while the ASR model learns the speech of the unknown vocabulary and the new text sequences not included in the paired data.

However, the conventional training that simultaneously trains both models with just the cycle-consistency loss causes the TTS model to over-adapt to the ASR model, which leads to two problems.
First, the synthesized speech after training does not reflect the speaker characteristics of the reference speech.
We assume this is because the synthesized speech is trained to have constant speaker characteristics that the ASR model most easily recognizes.
Second, pronunciation and prosody of the synthesized speech changes so that the ASR model easily recognizes them, which results in unnatural sounds for a human.
These two problems not only decreases the TTS model quality but also limits the ASR model improvement, as most of the synthesized speech becomes easy examples.

To solve these problems, we propose improving the cycle-consistency-based training with a speaker consistency loss and step-wise optimization.
The speaker consistency loss brings the speaker characteristics of the synthesized speech closer to that of the reference speech.
Specifically, it calculates the cosine similarity between the speaker embeddings of the synthesized speech and the reference speech, which has been used to train the multi-speaker TTS model in a supervised manner~\cite{ZCai20_SVtoMultiTTS,ATjandra20_MachineSpeechChain}.
We propose to use this loss during the semi-supervised joint training to preserve the speaker characteristics of the synthesized speech.
In the step-wise optimization, we freeze the parameter of the TTS model for a certain period at the beginning of the semi-supervised joint training.
This prevents the TTS model from over-adapting to the ASR model.
Our experiment shows that our proposed method achieves higher performance of both the ASR model and the TTS model compared to the conventional method.

\section{Preliminaries}
We utilize a Transformer-based ASR model~\cite{LDong18_TransformerASR} and a FastSpeech2-based TTS model~\cite{YRen21_Fastspeech2} in this work.
In this section, we describe the two models and explain how they are used in the conventional semi-supervised joint training.

\subsection{Transformer-based ASR model}
We denote the acoustic feature and its text as $\bm X = (\bm x_1, \ldots, \bm x_T)$ and $\bm y= (y_1, \ldots, y_L)$, respectively, where $\bm x_t\in\mathbb{R}^{F}$ denotes the $t$th frame of the feature, $F$ denotes its dimension, $T$ denotes the frame length, $y_l$ denotes the $l$th output token, and $L$ denotes the output sequence length.
In this paper, we use the time-frequency representation of the speech as the acoustic feature and the text at phoneme level.
The Transformer-based ASR model is trained to maximize the following posterior:
\begin{align}
  P(\bm y| \bm X; \bm \Theta_{\mathrm{asr}}) = \prod_l P(y_l|\bm y_{1:l-1}, \bm X; \bm \Theta_{\mathrm{asr}}),
\end{align}
where $\bm y_{1:l-1}=(y_1, \ldots, y_{l-1})$ and $\bm \Theta_{\mathrm{asr}}$ denotes the ASR model parameter.
The posterior is obtained by the encoder-decoder mechanism as follows:
\begin{align}
\bm H &= \mathrm{TransformerEnc}(\bm X;\bm \theta_{\mathrm{asr}}^{\mathrm{enc}}),\label{eq:asr1}\\
\bm c_l &= \mathrm{TransformerDec}(\bm H, \bm y_{1:l-1};\bm \theta_{\mathrm{asr}}^{\mathrm{dec}}),\label{eq:asr2}\\
\bm P(y_l&|\bm y_{1:l-1}, \bm X; \bm \Theta_{\mathrm{asr}}) = \mathrm{softmax}(\bm c_l;\bm \theta_{\mathrm{asr}}^{\mathrm{linear}})[y_l],\label{eq:asr3}
\end{align}
where
$\mathrm{TransformeEnc}(\cdot)$ is a transformer encoder that consists of a positional encoding layer and multiple multi-head self-attention blocks, $\bm \theta_{\mathrm{asr}}^{\mathrm{enc}}$ denotes its parameters,
$\mathrm{TransformeDec}(\cdot)$ is a transformer decoder that consists of an embedding layer, a positional encoding layer, and multiple multi-head self-attention and encoder-decoder attention blocks, $\bm \theta_{\mathrm{asr}}^{\mathrm{dec}}$ denotes its parameters,  $\mathrm{softmax}(\cdot)$ is a softmax layer with linear transformation, $\bm \theta_{\mathrm{asr}}^{\mathrm{linear}}$ denotes its parameters, and $[y_l]$ denotes the  element of the vector that corresponds to the probability of $y_l$.
The parameter $\bm {\Theta}_{\mathrm{asr}}=\{\bm \theta_{\mathrm{asr}}^{\mathrm{enc}},\bm \theta_{\mathrm{asr}}^{\mathrm{dec}},\bm \theta_{\mathrm{asr}}^{\mathrm{linear}}\}$ is optimized with cross-entropy loss function $L_{\mathrm{CE}}$ that is defined as
\begin{align}
L_{\mathrm{CE}} = -\frac{1}{L}\sum_l \log P(y_l|\bm y_{1:l-1},\bm X;\bm \Theta_{\mathrm{asr}}).\label{eq:asropt}
\end{align}
\subsection{FastSpeech2-based TTS model}
We design our multi-speaker TTS model based on FastSpeech2~\cite{YRen21_Fastspeech2} to improve training speed of TTS-ASR pipeline compared to autoregressive models such as Tacotron2~\cite{JShen2018_Tacotron2} and TransformerTTS~\cite{NLi2019_TransformerTTS}.
A reference speech is used to identify the speaker characteristics of the synthesized speech.
We denote the reference speaker's speech as $\tilde{\bm X}\in\mathbb{R}^{T'\times F}$, where $T'$ denotes the frame length of the reference speech.
The estimate of the acoustic features $\hat{\bm X}$ is obtained as follows:
\begin{align}
  \tilde{\bm s} &= \mathrm{SpeakerModel}(\tilde{\bm X};\bm \theta_{\mathrm{speaker}}),\label{eq:fs1}\\
  \bm U &= \mathrm{FastSpeech2Enc}(\bm y;\bm \theta_{\mathrm{tts}}^{\mathrm{enc}}),\label{eq:fs2}\\
  \bm V, \hat{\bm p}, \hat{\bm e}, \hat{\bm d}  &= \mathrm{VarianceAdaptor}(\bm U, \tilde{\bm s};\bm \theta_{\mathrm{tts}}^{\mathrm{va}}),\label{eq:fs3}\\
  \hat{\bm X}_{\mathrm D} &= \mathrm{FastSpeech2Dec}(\bm V;\bm \theta_{\mathrm{tts}}^{\mathrm{dec}}),\label{eq:fs4}\\
  \hat{\bm X} &= \mathrm{PostNet}(\hat{\bm X}_{\mathrm D};\bm \theta_{\mathrm{tts}}^{\mathrm{post}}),\label{eq:fs5}
\end{align}
where  $\mathrm{SpeakerModel}(\cdot)$ denotes the pretrained speaker model, $\bm \theta_{\mathrm{speaker}}$ denotes its parameters, $\tilde{\bm s}\in\mathbb{R}^{D_{\mathrm s}}$ denotes the speaker embedding, $D_{\mathrm s}$ denotes its dimension, $\mathrm{FastSpeech2Enc}(\cdot)$ denotes a FastSpeech2 encoder that consists of an embedding layer, a positional encoding layer, and multiple multi-head self-attention blocks, $\bm \theta_{\mathrm{tts}}^{\mathrm{enc}}$ denotes its parameters, $\mathrm{VarianceAdaptor}(\cdot)$ denotes a variance adaptor that consists of a pitch predictor, a energy predictor, a duration predictor, and a length regulator, $\bm \theta_{\mathrm{tts}}^{\mathrm{va}}$ denotes its parameters,
$\bm V$ denotes the phoneme hidden sequence with pitch, duration, and energy variation, $\hat{\bm p}\in\mathbb{R}^{L}$,  $\hat{\bm e}\in\mathbb{R}^{L}$, and $\hat{\bm d}\in\mathbb{R}^{L}$
are phoneme-wise pitch, energy, and duration, respectively, $\mathrm{FastSpeech2Dec}(\cdot)$ denotes a FastSpeech2 decoder that consists of a positional encoding layer and multiple multi-head self-attention blocks, $\bm \theta_{\mathrm{tts}}^{\mathrm{dec}}$ denotes its parameters, $\mathrm{PostNet}(\cdot)$ denotes a FastSpeech2 post-net, and $\bm \theta_{\mathrm{tts}}^{\mathrm{post}}$ denotes its parameters.
The TTS model parameter $\bm{\Theta}_{\mathrm{tts}}=\{\bm \theta_{\mathrm{tts}}^{\mathrm{enc}}, \bm \theta_{\mathrm{tts}}^{\mathrm{va}}, \bm \theta_{\mathrm{tts}}^{\mathrm{dec}}, \bm \theta_{\mathrm{tts}}^{\mathrm{post}}\}$ is optimized with the TTS loss that is defined as follows:
\begin{align}
  L_{\mathrm{TTS}} &= \|\bm X - \hat{\bm X}\|_1 +\|\bm X - \hat{\bm X}_{\mathrm D}\|_1\nonumber\\
  &\phantom{=}+\|\bm p - \hat{\bm p}\|_2^2 + \|\bm e - \hat{\bm e}\|_2^2 + \|\bm d - \hat{\bm d}\|_2^2, \label{eq:ttsopt}
\end{align}
where $\|\cdot\|_1$ denotes the L1 norm, $\|\cdot\|_2$ denotes the L2 norm, and $\bm p$, $\bm e$, and $\bm d$ denote the ground-truth pitch, energy, and duration, respectively.

\subsection{Conventional semi-supervised joint training}
In this section, we describe the baseline framework for semi-supervised joint training of TTS and ASR using paired data and unpaired text data.
First, the TTS model and the ASR model are separately pretrained with paired data using  \eqref{eq:asropt} and \eqref{eq:ttsopt}.
Then, for cycle-consistency training utilizing unpaired text data and paired data, a TTS-ASR pipeline is created, in which the multi-speaker TTS model synthesizes speech from unpaired text data using a random reference speech sampled from the paired data and the ASR model transcribes the synthesized speech to reconstruct the source text~\cite{ATjandra20_MachineSpeechChain,SKarita19_SharedTTSASR,THori19_CycleConsistencyTraining,YRen19_almostunsupervised,JXu20_LRSpeech,MBaskar21_EAT}.
The TTS-ASR pipeline is optimized with the cycle-consistency loss that is defined as
\begin{align}
  L_{\mathrm{cycle}} &= -\frac{1}{L}\sum_l \log P(y_l| \bm y_{1:l-1},\hat{\bm X};\bm{\Theta}_{\mathrm{asr}})\label{eq:ccloss},\\
  \hat{\bm X} &= \mathrm{TTS}(\bm y, \tilde{\bm X};\bm \Theta_{\mathrm{tts}}),\label{eq:tts_s}
\end{align}
where $\mathrm{TTS}(\cdot)$ is a function that combines Eqs. \eqref{eq:fs1}--\eqref{eq:fs5}.
Note that we omit $\hat{\bm p}$, $\hat{\bm e}$, and $\hat{\bm d}$ because they are not used in the semi-supervised joint training.
\section{Proposed method}
\begin{figure}
  \begin{center}
  \includegraphics[width=1.0\columnwidth]{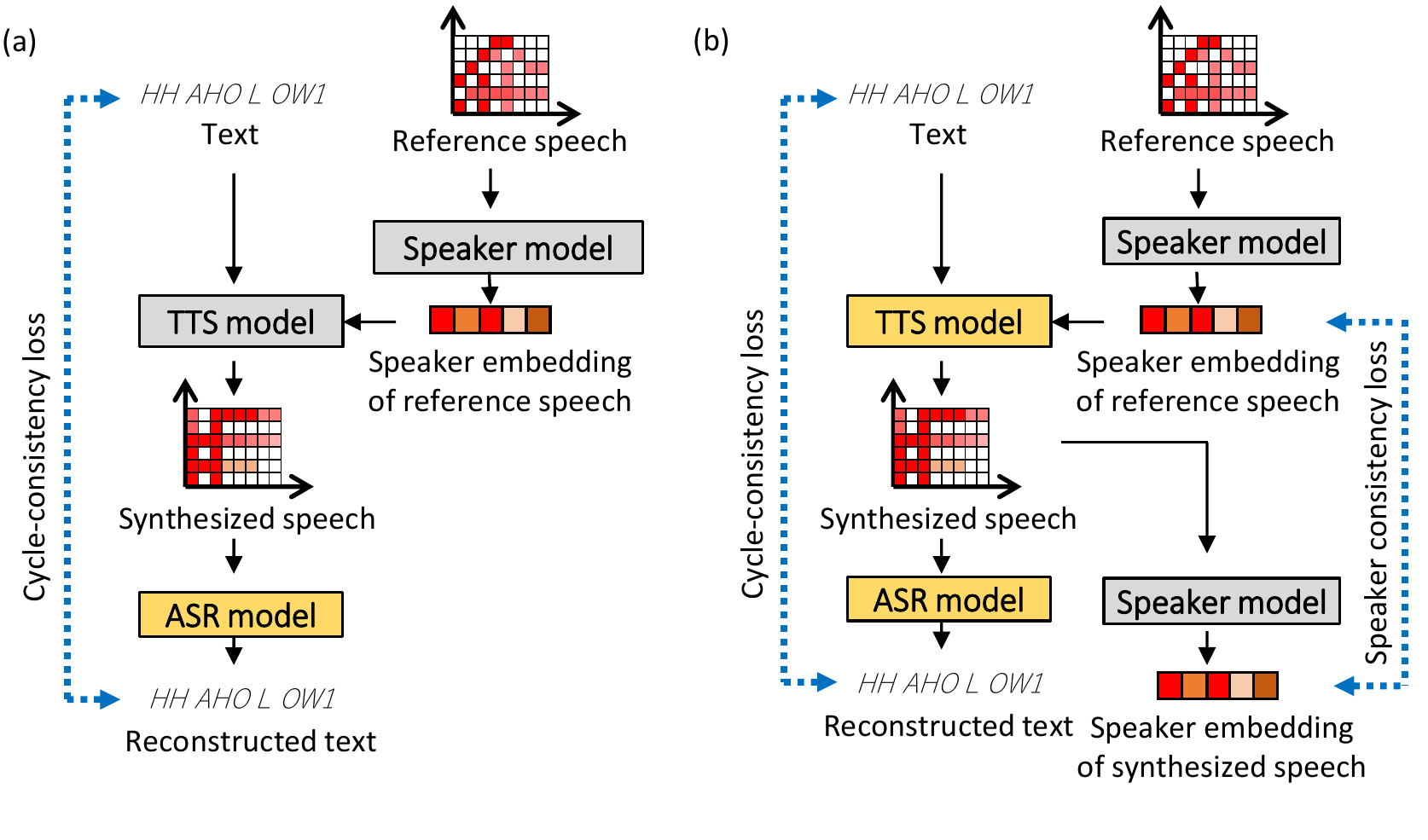}
  \vspace{-35pt}
  \end{center}
  \caption{Overview of proposed method in (a) first-step training and (b) second-step training.
  Orange and gray blocks represent updated and frozen models, respectively.}
  \vspace{-15pt}
  \label{fig:overview}
\end{figure}
\subsection{Strategy}
In conventional semi-supervised training utilizing unpaired text data, the TTS model and the ASR model are jointly trained with cycle-consistency loss \eqref{eq:ccloss}.
However, as discussed in section~\ref{sec:intro}, the speaker characteristics of the synthesized speech is not preserved in this case and the synthesized speech becomes overly easy for the ASR model to recognize, which leads not only to a low-quality TTS model but also to a limitation of the ASR improvement.
Our proposed method addresses this problem by utilizing step-wise optimization and the speaker consistency loss.

The overview of the proposed method is shown in Fig.~\ref{fig:overview}.
Our proposed step-wise optimization trains the ASR model and the TTS model in two steps.
In the first step (Fig.~\ref{fig:overview}(a)), we fix the parameter of the TTS model and optimize the ASR model.
In the second step (Fig.~\ref{fig:overview}(b)), we train both the TTS model and the ASR model.
The key point of this step-wise optimization is that the ASR model is optimized with the synthesized speech before the TTS model and ASR model are trained.
When the ASR model is not trained with the synthesized speech, the cycle-consistency loss becomes large at the beginning of the semi-supervised joint training, which drives the TTS model to synthesize speech that is easy for the ASR model to recognize.
In contrast, since the ASR model is trained with the synthesized speech after the first step of the proposed step-wise optimization, over-adaptation of the TTS model to the ASR model is prevented.
We experimentally show it in section~\ref{sec:result}.

The speaker consistency loss calculates the cosine similarity between the embeddings of the synthesized speech and that of the reference speech (shown in Fig.~\ref{fig:overview} (b)), and brings the speaker characteristics of the synthesized speech close to those of the reference speech.
This loss is beneficial for keeping the speaker characteristics of the synthesized speech.

\subsection{Training}
In this section, we formulate the speaker consistency loss and describe the step-wise optimization.
Given the synthesized speech $\hat{\bm X}$, we estimate its speaker embedding as
\begin{align}
  \hat{\bm s} &= \mathrm{SpeakerModel}(\hat{\bm X};\bm \theta_{\mathrm{speaker}}),\label{eq:prop1}
\end{align}
where the speaker model and its parameters are the same as those in \eqref{eq:fs1}.
We define the speaker consistency loss $L_{\mathrm{sc}}$ as
\begin{align}
  L_{\mathrm{sc}} = -\frac{\hat{\bm s}^{\mathrm T}\tilde{\bm s}}{\|\hat{\bm s}\|_2\|\tilde{\bm s}\|_2},
\end{align}
where $^\mathrm{T}$ denotes the transpose of a vector.
Minimizing the speaker consistency loss maximizes the similarity between embedding of the synthesized speech and that of the reference speech, which preserves the speaker characteristics of the synthesized speech.

We formulate the step-wise optimization with the speaker consistency loss as follows.
First, we freeze the parameter of the TTS model in the TTS-ASR pipeline and train the ASR model with the cycle-consistency loss ~\eqref{eq:ccloss}.
Then, we unfreeze the parameter of the TTS model and train both the ASR model and the TTS model with the cycle-consistency loss and the speaker consistency loss, which is defined as
\begin{align}
  L_{\mathrm{prop}}&= L_{\mathrm{cycle}}+\alpha L_{\mathrm{sc}}\label{eq:prop_all},
\end{align}
where $\alpha$ is the loss weight.
\section{Experiment\label{sec:exp}}
\subsection{Dataset}
We used the VoxCeleb2 dataset~\cite{JChung2018_Voxceleb2} for the speaker model training and the LibriTTS dataset~\cite{HZen2019_LibriTTS} for the TTS model and ASR model training and evaluation.
The speaker model was pretrained with a speaker classification of 5,994 speakers using the \texttt{dev} set of VoxCeleb2.
The \texttt{train-clean-100} set (train-100) and the \texttt{train-clean-360} set (train-360)  of LibriTTS  was used as paired data and unpaired text data, respectively.
We used \texttt{dev-clean} set and \texttt{test-clean-100} set of LibriTTS as validation data and test data, respectively.
We used 80 log mel-scale filterbank coefficients as acoustic features, which were extracted using a 50-ms-long Hann window with a 12.5-ms-long shift.
All the text was converted to phonemes and we inserted a word space token between each word.
To train FastSpeech2, we estimated the ground-truth duration, pitch, and energy of paired data with the Montreal Forced Aligner~\cite{MMcAuliffe17_mfa}, PyWorld~\cite{MMorise15_world1,MMorise15_world2}, and L2 norm of each frame following \cite{CChien21_Fastspeechprerocess}.
We used speech data with less than 900 frames and text data with less than 180 tokens for the memory constraint.
The sampling rate of all data was 16 kHz.

\subsection{Implementation}
The speaker model consisted of three bidirectional LSTM (BLSTM) layers with 256 units.
The forward and backward outputs of the BLSTM were concatenated.
The third BLSTM output was averaged with an attention mechanism to obtain the target speaker embedding, which calculated the weighted mean of frame-level features~\cite{KOkabe2018_Attentivepooling}.
The TTS model consisted of four encoder layers and six decoder layers.
We performed the phoneme-level prediction of energy and pitch as in \cite{CChien21_Fastspeechprerocess} and added the speaker embedding from the speaker model to the encoder outputs.
The post-net was the same as that of Tacotron2~\cite{JShen2018_Tacotron2}.
The other settings were the same as that in \cite{YRen21_Fastspeech2}.
The ASR model consisted of six encoder layers and four decoder layers.
The dimension of each transformer block was set to 512 and the number of attention heads was four.
\begin{table*}[t!]
  \caption{Evaluation results}
    \vspace{-20pt}
  \begin{center}
\begin{tabular}{|lwc{10mm}wc{18mm}lwc{7mm}wc{7mm}wc{12mm}|} \hline
  \textbf{Method} & \begin{tabular}{c}\textbf{Speaker}\\\textbf{consistency}\end{tabular} & \begin{tabular}{c}\textbf{Step-wise}\\\textbf{optimization}\end{tabular}&\textbf{Dataset}  &\begin{tabular}{c}\textbf{PER}\\\textbf{(\%)}\end{tabular} & \begin{tabular}{c}\textbf{MCD}\\\textbf{(dB)}\end{tabular} &\begin{tabular}{c}\textbf{F0}\\ \textbf{RMSE (Hz)}\end{tabular} \\ \hline

  Pretrained model & & &Paired train-100&15.5 & 8.3 & 36.7 \\
  Conventional method & & &Paired train-100 \& unpaired train-360&13.0 & 7.2 & 43.1 \\\hline
  Proposed method & \checkmark& \checkmark&Paired train-100 \& unpaired train-360& \textbf{9.6} & \textbf{6.9} & \textbf{33.2}\\
  \quad w/o speaker consistency & &\checkmark &Paired train-100 \& unpaired train-360&9.6 & 7.5 & 39.8 \\
  \quad w/o step-wise optimization &\checkmark & &Paired train-100 \& unpaired train-360& 12.5 & 6.9 & 34.8\\
  \hline\hline
  Full-supervised model & & &Paired train-100 \& paired train-360&6.6 & 7.7 & 37.0 \\
  \hline
\end{tabular}
\label{table:res}
\end{center}
\vspace{-20pt}
\end{table*}

\subsection{Settings}
We compared the following methods (listed in Table~\ref{table:res}): pretrained model trained with paired train-100, conventional method trained with paired train-100 and unpaired train-360, the proposed method trained with paired train-100 and unpaired train-360, and full-supervised model trained with paired train-100 and paired train-360.
To determine the effect of the speaker consistency loss and step-wise optimization, we also conducted an ablation study on each method.
The speaker model was optimized using the Adam~\cite{DKingma2015_Adam} algorithm with a minibatch size of 128.
We set the learning rate of the algorithm to $1e-3$.
The training steps were stopped if the loss on the validation set did not decrease for five epochs in succession.
We fixed the parameter of the speaker model after the pretraining.
The TTS model was optimized in the same way as the speaker model except that we set the minibatch size to 32 and the learning rate to $1e-4$.
The ASR model was optimized using the RAdam~\cite{LLiu20_Radam} algorithm with a minibatch size of 32.
The learning rate of the algorithm was set to $1e-4$.
Early stopping was used in the same way as the speaker model.
We used scheduled sampling~\cite{SBengio15_scheduledsample}-based optimization during the ASR training, where teacher forcing is used at the beginning of training, and we linearly  increased the probability of sampling to the probability of 0.4 at 20 epochs.
We also applied the time-masking and frequency-masking of SpecAugment~\cite{DPark2019_SpecAugment}, where the number of time-masks and frequency-masks is both two and the masking width is randomly set between 0 and 100 for the time-masking and between 0 and 27 for the frequency-masking.

For the semi-supervised joint training, we used the RAdam algorithm with a minibatch size of 8.
The learning rate was set to $1e-5$.
We set $\alpha$ in \eqref{eq:prop_all} to 0.1.
We trained the ASR model until convergence in the first step of the step-wise optimization.
We mixed the supervised paired data from train-100 and the unsupervised text data from train-360 and provided the mixed data randomly to the models.
When unsupervised text data was given, the models were optimized using \eqref{eq:ccloss} or \eqref{eq:prop_all}.
Otherwise, the models were optimized using \eqref{eq:asropt} and \eqref{eq:ttsopt}.
We applied the time-masking and frequency-masking of SpecAugment  to the synthesized speech in the same way as those during ASR pretraining.
The reference speech for the TTS model was randomly chosen from the paired data.
Since the parameter of the duration predictor of the TTS model was not differentiable, we fixed this parameter during training.
We used phoneme error rate (PER) as an ASR evaluation metric, and mel-cepstral distortion (MCD)~\cite{KRobert93_MCD} and root mean square error (RMSE) of fundamental frequency (F0) as the TTS evaluation metrics.
Dynamic time warping was used to calculate MCD and F0 RMSE.
We used MelGAN~\cite{KKumar19_MelGAN} as a vocoder and calculated F0 with DIO and StoneMask of PyWorld~\cite{MMorise15_world1,MMorise15_world2}.
To clarify the effect of the proposed step-wise optimization, we plotted the perplexity of the synthesized speech trained with and without step-wise optimization using validation data, which is calculated as
\begin{align}
  \mathrm{perp.} =\exp\left(-\frac{1}{L}\sum_l\mathrm{ln} P(y_l|\bm y_{1:l-1},\hat{\bm X};\bm{\Theta}_{\mathrm{asr}}^{\mathrm{pre}})\right),\label{eq:perp}
\end{align}
where $\bm{\Theta}_{\mathrm{asr}}^{\mathrm{pre}}$ denotes the parameter of the pretrained ASR model that is not trained with the synthesized speech.
The low perplexity indicates that the ASR model accurately estimates the posterior from the synthesized speech.
\begin{figure}[t]
  \begin{center}
  \includegraphics[width=0.6\columnwidth]{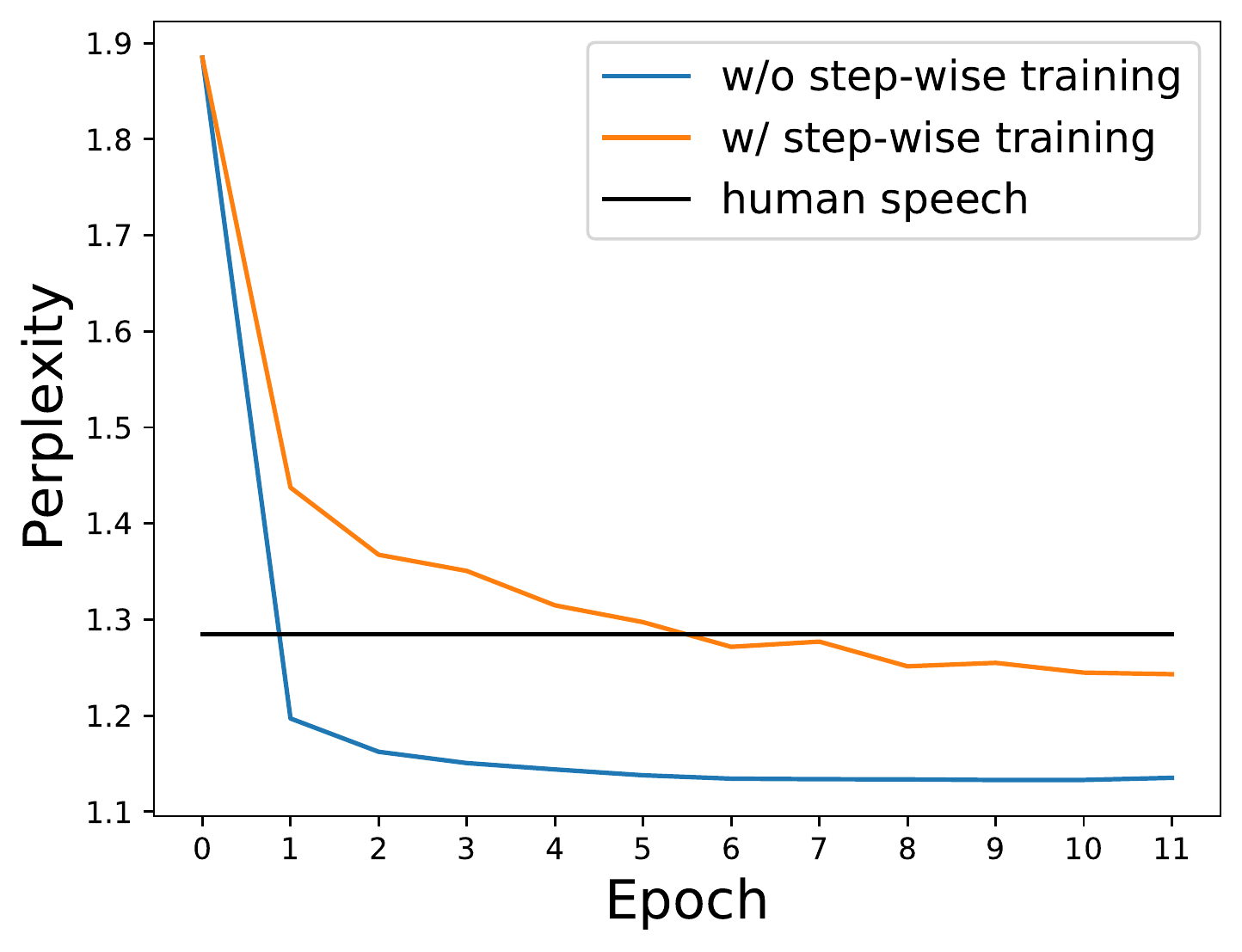}
  \vspace{-20pt}
  \end{center}
  \caption{Perplexity curve calculated using synthesized speech and human speech.
  Blue and orange lines represent perplexity calculated using synthesized speech trained without and with step-wise optimization, respectively. Black line represents perplexity calculated using human speech.}
  \vspace{-15pt}
  \label{fig:perp}
\end{figure}
\begin{figure}[t]
  \begin{center}
  \includegraphics[width=0.77\columnwidth]{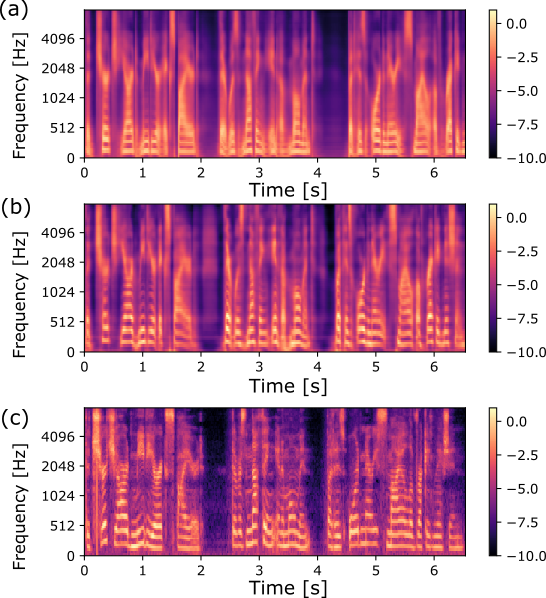}
  \vspace{-17pt}
  \end{center}
  \caption{Mel-spectrograms of (a) conventional method, (b) proposed method, and (c) ground truth.}
  \vspace{-18pt}
  \label{fig:synth}
\end{figure}

\subsection{Results\label{sec:result}}
Table~\ref{table:res} shows the result of each method.
As we can see, the proposed method outperforms the conventional method in terms of both ASR and TTS performance.
Specifically, the F0 RMSE of the conventional method is worse compared to that of the pretrained model.
This is because the semi-supervised joint training with just the cycle-consistency loss drives the TTS model to synthesize speech of the speaker that the ASR most easily recognizes.
In contrast, both MCD an F0 RMSE of the proposed method improve over the conventional method thanks to introducing the speaker consistency loss.
Table~\ref{table:res} also shows that the PER of the ASR model improves with our proposed step-wise optimization.
Figure~\ref{fig:perp} shows the perplexity curve of the proposed method with and without step-wise optimization.
We also plotted the perplexity calculated with the human speech as reference.
We can see here that the perplexity of the TTS model trained without step-wise optimization becomes far lower compared to human speech after just one epoch, which indicates that the TTS model over-adapts to the ASR model and the synthesized speech becomes overly easy for the ASR model to recognize.
In contrast, the perplexity curve of the TTS model trained with the proposed step-wise optimization becomes gentle, and perplexity converges to the point closer to that calculated with human speech, which indicates the proposed step-wise optimization prevents the over-adaptation of the TTS model to the ASR model.
Figure~\ref{fig:synth} shows the mel-spectrograms of the conventional method, the proposed method, and the ground truth.
The proposed method reflected the characteristics of the ground-truth speech that the conventional method failed to reflect.

\section{Conclusion}
In this paper, we proposed to improve the cycle-consistency-based training with the speaker consistency loss and step-wise optimization.
The speaker consistency loss brings the speaker characteristics of the synthesized speech closer to that of the reference speech.
The step-wise optimization prevents the over-adaptation of the TTS model.
Experimental results showed that the speaker consistency loss improves TTS quality in terms of MCD and F0 RMSE, and the step-wise optimization prevents over-adaptation of the TTS model, which leads to the improvement of both the ASR model and the TTS model.

\end{document}